\documentclass{article}
\usepackage{ijcai19}

\usepackage{times}
\usepackage{soul}
\usepackage{url}
\usepackage[hidelinks]{hyperref}
\usepackage[utf8]{inputenc}
\usepackage[small]{caption}
\usepackage{graphicx}
\usepackage{amsmath}
\usepackage{booktabs}
\newcommand{\strategy}{\Sigma}

\title{The Anatomy of Corner 3s in the NBA: What makes them efficient, how are they generated and how can defenses respond?}

\author{
Konstantinos Pelechrinis$^1$\and
Kirk Goldsberry$^2$
\affiliations
$^1$Department of Informatics and Networked Systems, University of Pittsburgh\\
$^2$McCombs School of Business, University of Texas, Austin\\
}

\begin{document}

\maketitle

\begin{abstract}
Modern basketball is all about creating efficient shots, that is, shots with high payoff. 
This is not necessarily equivalent to creating looks with the highest probability of success. 
In particular, the two most efficient shots in the NBA - which are shots from the paint, i.e., extremely close to the basket, and three-point shots from the corner, i.e., at least 22 feet apart - have completely different spatial profiles when it comes to their distance from the basket. 
The latter also means that they are pretty much at the opposing ends of the spectrum when it comes to their probability of being made. 
Due to their efficiency, these are the most sought after shots from the offense, while the defense is trying to contain them. 
However, in order to contain them one needs to first understand what makes them efficient in the first place and how they are generated. 
In this study we focus on the corner three point shots and using player tracking data we show that the main factor for their efficiency - contrary to the belief from the sports mass media - is not the shorter distance to the basket compared to three-point shots above the break, but rather the fact that they are assisted at a very high rate (more than 90\%). 
Furthermore, we analyze the movement of the shooter and his defender and find that more than half of these shots involve a shooter {\em anchored} at the corner waiting for the kick out pass. 
We finally define a simplified game between the offense and defense in these situation and we find that the Nash Equilibrium supports either committing to the corner shooter or to the drive to the basket, and not lingering between the two, which is what we observed from the defenses in our dataset. 
\end{abstract}

\section{Introduction}

\begin{figure*}
    \centering
    \includegraphics{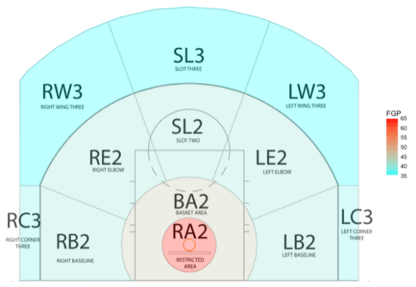}
\includegraphics{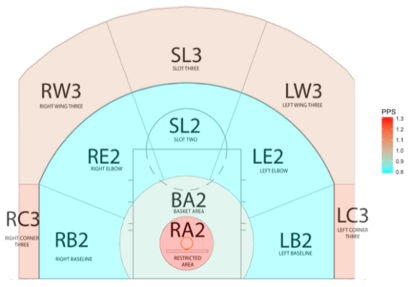}
    \caption{While FG\% monotonically reduces with distance (left figure), corner three point shots are second only to shots in the restricted area in terms of efficiency (right figure). }
    \label{fig:heatmaps}
\end{figure*}

For many years media and fans have been {\em evaluating} a shot based on the chance of it going through the basket. 
A shot with a 44\% chance of being made, should be better than a shot that has a 38\% chance of being made. 
Given that the chance of making a shot ceteris paribus is monotonically decreasing with distance, shots further away from the basket would be deemed of lower quality. 
This would be correct if all shots were worth the same number of points. 
However, given that a team wins by scoring points and not by having a better field goal percentage (FG\%), taking a 22-feet three point shot from the corner provides an expected point value of 1.14 and is more {\em valuable} as compared to a 20-feet jumper that is worth 2 points that provides an expected value of 0.9 points.  
Figure \ref{fig:heatmaps} depicts the two different views on shot value (chance of making the shot - vs - expected point value from the shot). 

From this figure we can see that while shots from the restricted area are the most efficient ones, three-points shots from the corners are a close second. 
Midrange shots on the other hand are less efficient than any type of three-point shot, despite the fact that they are being made at a higher rate. 
These - basic - realizations are really the root of the well-documented trend for the increased volume in three-point attempts in the past decade \cite{goldsberry2019sprawlball}. 
However, even among the three point shots there are differences in efficiency among shots taken from the corner (C3 for short) and those taken from above the break (ATB3 for short). 
In particular, shots from the corners generate about 12 points per 100 shots more as compared to ATB three-point shots. 
This has made C3s particularly appealing to offenses and even led to the so-called ``corner specialist'' player type \cite{goldsberry2019sprawlball}. 

Nevertheless, there are still important aspects of this shot that we do not have a good understanding of. 
For example, when media, fans and even professionals \cite{breport} talk about its efficiency, they cite the shortest distance from C3 as compared to ATB3. 
While this is a contributing factor, it is not the sole or even the main reason that drives C3's efficiency. 
In this study, we focus on answering the following three questions: 

\begin{itemize}
    \item What makes C3s so efficient, and more specifically, why are they 12 points/100 shots better than ATB3?
    \item How do offenses generate these shots? 
    \item How are defenses attempting to defend C3s and what is the {\em best} defensive approach?
\end{itemize}

We answer these questions by using a variety of different data, including player tracking data from the NBA, as well as, shot data from an international FIBA competition. 
Our main findings can be summarized in the following: 

\begin{itemize}
    \item While the shortest distance from the corners improves the FG\% of a C3 shot, the gap in the efficiency over ATB3s is best explained considering the fact that they are assisted at a very high rate ($>90\%$). This results in them being less contested as compared to ATB3. 
    \item Approximately half of all the C3s in our data, involved a shooter who was anchored at the corner for at least 4 seconds prior to the shot and received a kick out pass. 
    \item Defenses do not commit to the C3 shooter and stay approximately 10 feet away from them - between the corner and the basket, in an attempt to provide help defense for the drive. 
    \item We solve a simplified game capturing this situation and find that the Nash equilibrium dictates that the defender either commits to the corner shooter or to help defense at the basket. The exact mixed strategy depends on the shooting and driving abilities of the players involved. 
\end{itemize}

The rest of the paper is organized as follows: 
In Section \ref{sec:related} provides the relevant background, while it also describes the data used in the study. 
Section \ref{sec:analysis} presents in detail our analysis and the results. 
Finally, Section \ref{sec:conclusions} concludes our study by discussing its limitations and future directions. 

\section{Background and Experimental Setup}
\label{sec:related}

{\bf Player Tracking Data and Defensive Metrics: }The availability of player tracking data has allowed researchers and practitioners in sports analytics to analyze and model aspects of the game that were not possible with traditional data at a large scale. 
For example, defensive metrics have traditionally been limited to measuring steals and blocks, ignoring various complex defensive aspects such as off-ball defense, help defense, closeouts etc. 
Tracking data have allowed to slowly close the gap between what can be recorded in the boxscore and defensive value. 
\cite{franks2015counterpoints} developed models for capturing the defensive ability of players based on the spatial information obtained from optical tracking data. 
Their approach is based on a combination of spatial point processes, matrix factorization and hierarchical regression models and can reveal information that cannot be inferred with boxscore data. 
As an example, the proposed model can identify whether a defender is effective because he reduces the quality of a shot or because he reduces the frequency of the shots altogether. 
\cite{seidlbhostgusters} further used optical tracking data to learn how a defense is likely to react to a specific offensive set using reinforcement learning, while \cite{chen2018generating} utilize conditional generative adverserial networks for the same task. 
These models can further be combined with an expected points model \cite{cervone2016multiresolution,sicilia2019deephoops} to evaluate the defensive performance of a team/player based on the differences of the expected outcome between the observed defensive reaction and the {\em league average}.  

{\bf Game Theory and Basketball: } 
Game theory has been used to analyze end-of-game strategy in basketball \cite{winston2012mathletics}. 
\cite{nc-gametheory} defined a simplified game to study the evolution of shooting and computes the Nash equilibrium for different situations. 
Furthermore, \cite{skinner2010price} utilized the notion of price of anarchy to show the plausibility of the so-called ``Ewing Theory'', that is, removing a key player from a team can result in the improvement of the team's offensive efficiency. 
\cite{spv21} used the notion of Shapley values from the theory of cooperative games to develop a player evaluation metric for single basketball games.

{\bf Datasets: }
To complete our study we use player tracking data from 750 games from the 2016-17 NBA season. 
These data, captured through six cameras placed on the rafters of the arenas, include information for the locations of the players and the ball sampled at a rate of 25Hz. 
We also use data from an international competition. 
In particular, we use shot data from the 324 games of the 2016-17 season of FIBA's Basketball Champions League Competition. 

\begin{figure*}
    \centering
    \includegraphics[scale=1.15]{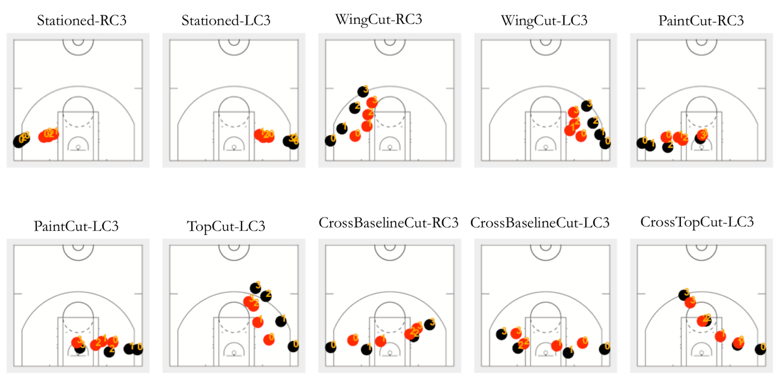}
    \caption{The 10 clusters identified for the shooter-defender movement before a C3 shot. The different points represent the shooter and his defender for the four seconds prior to the shot.}
    \label{fig:clusters}
\end{figure*}

\section{Analysis and Results}
\label{sec:analysis}

In this section we will present our analysis, where we answer the questions posited in the introduction. 

\subsection{Why are C3s more efficient than ATB3? }

As aforementioned, the common belief among media and analysts is that the shorter distance from basket is responsible for this difference in efficiency. 
However, as we will see this explains only part of the efficiency gap. 
In particular, we start by building a logistic regression model for the probability of a shot being made based solely on the distance between the shooter and the basket. 
According to this model, C3s are indeed expected to have a higher chance of being made, when compared to ATB3s. 
The expected difference is approximately 1.8\%, that is, we expect C3s to have a FG\% of about 1.8 percentage points higher than ATB3s. 
This also accounts for the fact that ATB3s can be much further away than the line, while there is little room for a C3 shot, by using the average distance of a ATB3 and C3 shot - 25.1 feet and 23 feet respectively - instead of the line distance. 
However, in our data the gap between the two shots is much higher. 
C3s are made at a 38.8\% rate, while ATBs exhibit a FG\% of 34.7\%; a difference of 4.1\%. 
Hence, the shorter distance explains only part of the observed difference in the efficiency between the two shots. 
What factors can explain the rest of the efficiency gap between the two types of three-point shots? 

For identifying this factor(s) we should think of what are some other variables that can alter the chance of a shot being made. 
One of the most intuitive variables is the level of contest from the defense. 
This can be captured - even though not perfectly - by the distance between the shooter and the closest defender. 
When calculating the distance between the shooter and the closest defender we found that C3 shots are {\em less contested} as compared to ATB3, with the closest defender being an average of 6.5 feet away at the time of the shot (while this distance is a bit lower than 6 feet for ATB3s). 
Nevertheless, this distance still does not capture the on-court mechanism that drives the efficiency. 
This mechanism is assists! 
Assisted shots are on average more open compared to unassisted shots, as they are typically the result of good ball movement that allows to find the open player. 
In our data, more than 90\% of the C3 shots are assisted, while ATB3s are assisted at a lower rate of just above 70\%. 

Of course, the above analysis cannot be treated as causal. 
Ideally, following the gold standard for causal inference, we would randomly assign half of the games to be played on a court where the distance from the three-point line is the same everywhere, and the rest assigned to be played on a regular NBA court. 
Nevertheless, as you can probably imagine this is not possible (even in the era of NBA bubble). 
However, there is a natural experiment that we can take advantage of to strengthen our above argument. 
In particular, the three-point line for FIBA competitions has different geometry, with the difference in the distance to the basket between C3 and ATB3 shots being only 28\% of that in the NBA court. 
If distance was the main factor driving the efficiency gap between C3 and ATB3 shots, this gap should significantly diminish, or even be eliminated altogether, in FIBA competitions. 
Using data from FIBA's Champions League competition we find that this is not the case. 
The efficiency gap between the two different types of shot is still present, with C3s providing 16 points/100 shots more compared to ATB3\footnote{The efficiency gap is higher than the one observed in the NBA but this could be due to the much smaller sample size we have from the FIBA competition.}. 
Furthermore, C3s in our FIBA dataset are still assisted at a much higher rate - approximately 45\% higher - than ATB3. 
Even though there are no player tracking data available for FIBA competitions, we expect the higher assist rate for C3s to also lead to less contested shots similar to the NBA. 

{\bf Takeaway: } The shorter distance to the basket explains only part of the efficiency gap betweeen C3s and ATB3s. 
The main driving factor appears to be the fact that C3s are a result of good ball movement resulting in assisted, less contested shots.

\subsection{How do offenses generate C3 shots?}

\begin{figure}
    \centering
    \includegraphics[scale=0.32]{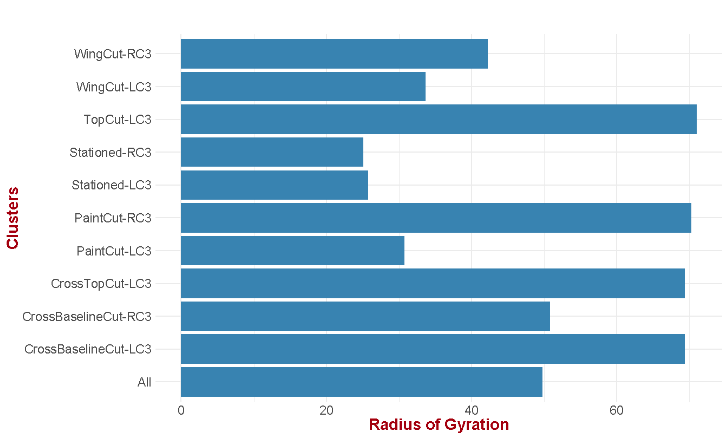}
    \caption{The two ``Stationed'' clusters exhibit the smaller variability in terms of shooter-defender movement.}
    \label{fig:gyration}
\end{figure}

Next we are interested into understanding better how offenses are able to generate these C3 shots. 
In particular, we are interested in the {\em shooter-defender choreography} for a time window before the shot is taken. 
We focused on the shooter and the defender and, in particular, their location for the last four seconds before the shot. 
Using k-means and examining a different number of clusters, we clustered the trajectories of the players into 10 clusters based on the gap statistic \cite{tibshirani2001estimating}. 
These clusters are shown in Figure \ref{fig:clusters}. 
The first two clusters (termed as ``Stationed-RC3'' and ``Stationed-LC3'') include half of the corner three shots in our data! 
In these instances, a player is (literally) anchored in the corner and waits for a drive-n-kick, while his defender is lingering between the basket (e.g., to double team the penetration) and the shooter. 
We further calculate the radius of gyration for each cluster:

\begin{equation}
    r_C=\sqrt{\dfrac{1}{||C||}\sum_{\mathbf{x}_i \in C}(\mathbf{x}_i-\mathbf{x}_C)^2}
\end{equation}
where $\mathbf{x}_i$ is a data point belonging in cluster $C$ and $\mathbf{x}_C$ is the centroid of the cluster. 
The radius of gyration measures the standard deviation of distances between the cluster members and the centroid of the cluster. 
Therefore, low radius of gyration corresponds to a spatially coherent cluster, i.e., all the instances are close to the centroid, which in turns is representative of the group. 
Figure \ref{fig:gyration} presents the radius of gyration for each cluster. 
As we can see, the ``Stationed'' clusters are not only the most common ones, but they also have the smallest variability. 

One important aspect missing in the above is the ball location and where the ball is delivered to the corner from. 
Table \ref{tab:pass} shows the fraction of passes delivered in the right and left corner from the different court zones (the court zones correspond to the ones shown in Figure \ref{fig:heatmaps}). 
As we can see almost one third of the passes come from the area around the basket (basket area, deep paint and baseline). 
In turns, these typically correspond to kick-out passes after a drive.  

\begin{table}[]
    \centering
    \begin{tabular}{c||c|c}
   {\bf Court Zone}     & {\bf Left Corner} & {\bf Right Corner} \\
    \hline
    Basket Area     & 20.6\% & 22.1\% \\
    Deep Paint & 5.8\% & 6.5\% \\
    Left Baseline & 3.7\% & 2.4\% \\
    Left Corner & 0\% & 0\% \\
    Left Wing 2 & 17.1\% & 12.3\% \\
    Left Wing 3 & 17.3\% & 2.9\% \\
    Midrange Slot & 13.3\% & 13.7\% \\
    Right Baseline & 1.8\%  & 2.4\% \\
    Right Corner & 0\% & 0\% \\
    Right Wing 2 & 10.7\% & 16.1\% \\ 
    Right Wing 3 & 3.3\%   & 15.3\% \\
    Top of Arc &  6.4\% & 6.3\% \\
    \end{tabular}
    \caption{The majority of the passes for C3 shots come from the paint (Basket Area and Deep Paint)}
    \label{tab:pass}
\end{table}

{\bf Takeaway: } 
Approximately half of the C3 shots include the shooter patiently waiting at the corner for the pass. 
Approximately a third of these passes come from near the basket. 

\subsection{How can defenses counter C3s? }

\begin{figure}
    \centering
    \includegraphics[scale=0.45]{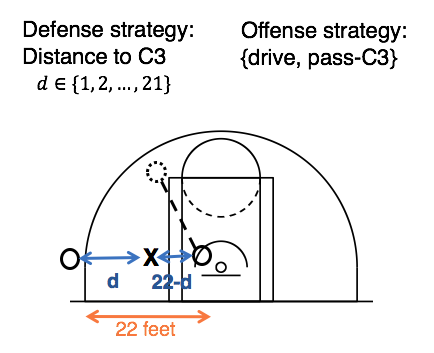}
    \caption{A simplified game for defending C3 shots. }
    \label{fig:game}
\end{figure}

In order to gain insights for this question we turn to game theory. 
We will define and analyze a simplified game, and while many details have been abstracted, this model can still provide good insights on how teams might want to alter their defensive strategy.  
Given that a large portion of the C3 shots involve a drive and kick-out pass to a stationed shooter at the corner we will define a game that focuses on this situation. 
Figure \ref{fig:game} depicts this simplified game. 
The main question is what should the defender assigned to the shooter do. 
He has two options; (i) commit to his assignment, (ii) attempt to double team the penetration. 
In reality, his strategy $\strategy_{def}$ is the distance $d$ he keeps from the shooter, which we quantize in bins of 1 foot, and hence, $\strategy_{def} = \{1, 2, 3, \dots, 21\}$.  
The offense has two strategies $\strategy_{off}$ in this situation; drive to the basket or make the kick-out pass to the shooter.  

\begin{figure}
    \centering
    \includegraphics[scale=0.35]{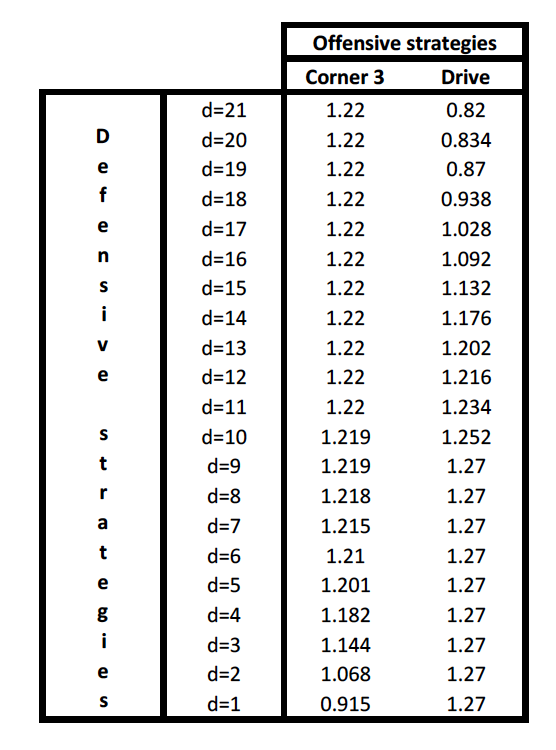}
    \caption{Payoff matrix for $\alpha = 2$}
    \label{fig:payoff_m}
\end{figure}

Given that the payoff for either player equals the loss of the other this is a zero-sum game. 
In order to solve this game, we need to first define the payoff matrix $\mathbf{\Pi}$. 
The payoff matrix (from the perspective of the offense) represents the expected points scored based on the corresponding strategies. 
Therefore, we need to identify the points per shot from the corner as a function of the distance of the closest defender. 
We can use Second Spectrum's qSQ \cite{chang2014quantifying} to identify the league-average points per shot as a function of the distance of the closest defender. 
We can also simulate the impact of a second defender in a penetration close out through a parametrized model. 
For example, we consider that a second defender {\em playing} strategy $d \in \strategy_{def}$, i.e., being at distance $d$ from the corner (and hence $22-d$ from the player penetrating) will reduce the expected points per shot of the drive by a factor of $(1-\dfrac{1}{\alpha^{(22-d)}})$. 
Different values of $\alpha$ correspond to different levels of impact on a double team. 
For example, for $\alpha = 2$ the payoff matrix (for the offense) is presented in Figure \ref{fig:payoff_m}.

\begin{figure*}
    \centering
    \includegraphics[scale=0.4]{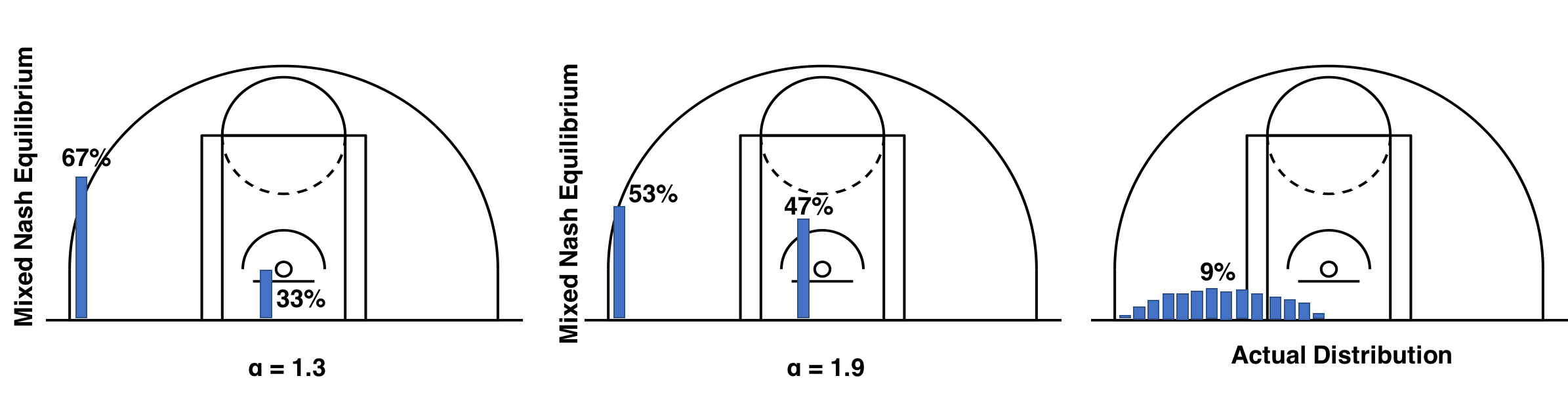}
    \caption{The Nash equilibrium of our simplified game suggests that the corner defender should commit to either helping the penetration or stay with their player at the corner.}
    \label{fig:nash}
\end{figure*}

We solved the game for several values of $\alpha$ between 1 and 2, and Figure \ref{fig:nash} shows the Nash equilibrium mixed strategy for the corner defender for two specific values ($\alpha = 1.3$ and $\alpha = 1.9$). 
As we can see, the Nash equilibrium strategy essentially tells us that the defender should commit to one of the two options (either guarding the corner or double teaming the penetration). 
Depending on the value of $\alpha$ (which controls the effectiveness of the double team), the fraction of time for each option changes. 
However, this simple game theory model says that one should not choose the ``in-between'' defenses.

An interesting point is that the expected distance $\hat{d}$ of the corner defender based on this Nash equilibrium will be close to 13 feet, with the exact value depending on the model parameters ($\alpha$). 
This value is close to the average of the observed distances of the defender from the corner three shooter, which is equal to 12.3 feet. 
However, as we can see in the third part of Figure \ref{fig:nash}, the distribution is very different than the Nash equilibrium mixed strategy. In fact, the empirical distribution of the distances resembles a normal distribution around the average instead of the bi-modal mixed Nash equilibrium strategy.

{\bf Model limitations: }
The game we created is based on a league-average shooting model. 
However, the payoff matrix can be adjusted to specific situations to account for the ability of the C3 shooter, the ability of a player to finish the drive, as well as, the defensive ability of the defender. 
It also does not currently account for things such as the ability to make a pass to the corner when double teamed in a penetration. 
Nevertheless, similar factors can be integrated in the payoff matrix. 
For example, if a player completes 80\% of his passes to the corner, we can adjust the payoff $\pi_{d,pass}^{new}=0.8\cdot \pi_{d,pass}$, where $\pi_{d,pass}$ essentially captures the expected points from the C3 (i.e., the payoff when assuming all passes are successful).  
Despite its limitations this model provides us with useful insights on how teams should rethink their defensive strategy. 
In fact, the 76ers - famed for their interest in incorporating analytics in their approach - have been reported to be cognizant of this. 
76ers' point guard Ben Simmons said: ``Numbers can only tell you so much, but they can really tell you how to guard certain teams. We know what other teams are good at and what they are not good at. We usually play to that. People sometimes think we’re stupid, but they don’t know what we know. To be fair, some of the information and actions are counterintuitive. For instance: when you play a team with deadly corner three-point shooters, it’s smarter to stay on your man in the corner and not help when an opponent drives down the lane, and that can look foolish for the unwashed'' \cite{76ers}.

{\bf Takeaway: }
The optimal defensive strategy involves committing to either guarding the corner shooter or double teaming the drive. 
The exact percentage of time spent on each strategy depends on contextual factors (quality of the shooter, quality of the player driving etc.), but the dominant strategy currently employed by NBA teams of placing the defender in between the basket and the corner appears to be suboptimal. 

\section{Discussion and Conclusions}
\label{sec:conclusions}

In this study we focused on corner three-point shots, one of the most efficient shots in modern basketball, second only to attempts at the rim. 
We start by showing that while the shorter distance at the corners explains part of the efficiency gap, the major driver of C3's efficiency is the fact that they are assisted. 
They are the result of (good) ball movement and spacing, leading them to be less contested as compared to ATB3 shots. 
By analyzing the movements of the corner shooter and his defender, we see that in more than half of C3 shots, the shooter is anchored at the corner waiting for a kick-out pass for at least 4 seconds. 
We finally find the Nash equilibrium of a simplified game that captures the interactions between the offense and defense in ``drive and kick'' situations at C3. 
The mixed Nash equilibrium indicates that the corner defender should commit to either guarding the corner shooter or doubling the drive. 
Current defenses deviate significantly from this, with the defender staying in between the corner and the basketball. 

\begin{figure*}
    \centering
    \includegraphics[scale=0.55]{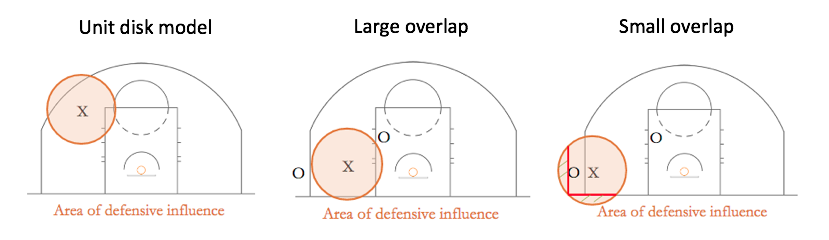}
    \caption{Assuming a mental representation of a unit disk model for the defensive influence of a player (left), lingering between the corner and the three maximizes the overlap between the disk and the court (middle), while committing to the corner reduces the overlap (right). }
    \label{fig:unitdisk}
\end{figure*}

{\bf Future work: }
The natural follow up question to be answered is why do defenders linger between the corner and the basket instead of committing to one of the two as the Nash equilibrium would dictate? 
While we do not have the answer to this question we put forth some plausible research hypothesis that we will examine in our future work. 
More specifically, it is plausible that this is a result of underlying cognitive biases associated with assessing spatial risk. 
A simplified mental model/representation for the spatial defensive influence can be that of a unit disk model. 
In a unit disk model, the defender exerts influence/disrupts in an area centered around him of radius $\rho$ (left part Figure \ref{fig:unitdisk}). 
Therefore, it comes natural to a defender to want to position himself in a way that maximizes the overlap of his unit disk with the court. 
The middle and right parts of Figure \ref{fig:unitdisk} further visualizes the situation, where as we see when the defender commits to the corner shooter part of his unit disk is outside of the court, possibly making him believe his influence on the defensive side is smaller as compared to when he is between the corner and the basket. 
Of course, the fallacy in this mental representation is that the areas on the court are not all of the same value and the intersection between the court and the defender's unit disk needs to be appropriately weighted by the court value. 

As part of our future work we will design appropriate lab experiments to examine - and refine if needed - this hypothesis. 
If it turns out to be true, it is important to identify possibly behavioral interventions that will allow coaching stuff to create the habits for the players that will allow them to make the appropriate decisions during games. 
Behavioral interventions are not something new for the NBA, as teams have explored a variety of approaches during practice to improve various aspects of their game such as spacing \cite{76ers-wsj,mavs-beh}.

{\bf Acknowledgments:} We would like to thank FIBA Basketball Champions League for providing us access to their data API.
 
 \bibliographystyle{named}
\bibliography{ijcai19}
 
\end{document}